\documentclass[prl, floatfix, twocolumn, showpacs, superscriptaddress]{revtex4-2} 

\usepackage{amsmath, amsfonts, amssymb, dsfont}
\usepackage{mathtools}
\usepackage[dvipsnames,table,xcdraw]{xcolor}  
\usepackage{float}
\usepackage[caption=false,font=footnotesize]{subfig}
\usepackage{graphicx}
\usepackage{tabularx}
\usepackage{xargs}  
\usepackage{soul} 
\usepackage[export]{adjustbox}
\usepackage{xprintlen} 
\usepackage{wasysym}
\usepackage{array}
\usepackage[toc,page,titletoc]{appendix}
\usepackage{pgfplots}
\usepackage{standalone}
\usepackage{import}
\usepackage[hypertexnames=false]{hyperref}
\usepackage[capitalize]{cleveref} 
\usepackage[normalem]{ulem}
\usepackage{subfiles}
\usepgfplotslibrary{groupplots}
\pgfplotsset{compat=newest}
\hypersetup{
   colorlinks,
   citecolor=blue,
   filecolor=blue,
   linkcolor=blue,
   urlcolor=blue,
   breaklinks=true
}

\newcommand*\ket[1]{\left\vert #1 \right\rangle}

\newcommand*\braket[2]{\left\langle #1 \vert #2 \right\rangle}

\newcommand*\SE{S_\mathrm{E}} 
\newcommand*\SN{S_\mathrm{N}} 
\newcommand*\VarX{\sigma^2_{x_n\mspace{-10mu}}~} 
\newcommand*\VarXInfty{{\sigma^2_{x_n\mspace{-10mu}}}^\infty} 
\newcommand*\davgSE{\overline S_\mathrm{E}} 
\newcommand*\davgSN{\overline S_\mathrm{N}} 
\newcommand*\davgVarX{\overline{\sigma^2_{x_n\mspace{-10mu}}}~} 
\newcommand*\davgVarXInfty{\overline{\sigma^2_{x_n\mspace{-10mu}}}^\infty} 
\newcommand*{\xiJtwo}{\xi_J^\prime} 
\newcommand*{\xiJthree}{\xi_J^{\prime\prime}} 

\newcommand*{\internalwhtsqr}{{\mathrlap{\raisebox{-0.3pt}{\scalebox{0.90}{\ensuremath\square}}}\hphantom{\rhd}}}
\newcommand*{\internalwhtcrc}{{\mathrlap{\ensuremath\Circle}\hphantom{\rhd}}}
\newcommand*{\internalwhtrhd}{{\mathrlap{\ensuremath\rhd}\hphantom{\rhd}}}
\newcommand*{\internalwhtlhd}{{\mathrlap{\ensuremath\lhd}\hphantom{\lhd}}}
\newcommand{\makemathchoices}[1]{{%
    \mathchoice%
    {#1}
    {#1}
    {\scalebox{0.75}{\ensuremath#1}}
    {\scalebox{0.5}{\ensuremath#1}}
}}

\newcommand*{\whtsqr}{\makemathchoices{\internalwhtsqr}}
\newcommand*{\whtcrc}{\makemathchoices{\internalwhtcrc}}
\newcommand*{\whtrhd}{\makemathchoices{\internalwhtrhd}}
\newcommand*{\whtlhd}{\makemathchoices{\internalwhtlhd}}
\newcommand*\csqr{\mathbin{\vcenter{\hbox{\rule{.3ex}{.3ex}}}}}
\newcolumntype{?}{!{\vrule width 1.5pt}}

\begin{document}
\title{Ultraslow Growth of Number Entropy in an \texorpdfstring{$\ell$-bit}{l-Bit} Model of Many-Body Localization}
\author{David Aceituno Chávez}
\affiliation{Department of Physics, KTH Royal Institute of Technology, Stockholm, 106 91 Sweden}
\author{Claudia Artiaco}
\affiliation{Department of Physics, KTH Royal Institute of Technology, Stockholm, 106 91 Sweden}
\author{Thomas Klein Kvorning}
\affiliation{Department of Physics, KTH Royal Institute of Technology, Stockholm, 106 91 Sweden}
\author{Loïc Herviou}
\affiliation{Univ. Grenoble Alpes, CNRS, LPMMC, 38000 Grenoble, France}
\affiliation{Institute of Physics, Ecole Polytechnique Fédérale de Lausanne (EPFL), CH-1015 Lausanne, Switzerland}
\author{Jens H. Bardarson}
\affiliation{Department of Physics, KTH Royal Institute of Technology, Stockholm, 106 91 Sweden}

\begin{abstract}
We demonstrate that slow growth of the number entropy following a quench from a local product state is consistent with many-body localization.
To do this, we construct a novel random circuit $\ell$-bit model with exponentially localized $\ell$-bits and exponentially decaying interactions between them.
We observe an ultraslow growth of the number entropy starting from a Néel state, saturating at a value that grows with system size. 
This suggests that the observation of such growth in microscopic models is not sufficient to rule out many-body localization.   
\end{abstract}

\maketitle

The many-body localized insulator~\cite{Basko.2006tfa,Gornyi.2005} is widely believed to be an emergent phase of interacting quasilocal integrals of motion, often referred to as $\ell$-bits~\cite{Serbyn.2013, Huse.2014, Abanin.20192a9}.
The $\ell$-bits in spin-$1/2$ models are dressed physical on-site Pauli spin operators~\cite{Serbyn.2013, Huse.2014}, while in fermionic models they are Anderson orbitals dressed by local electron-hole excitations~\cite{Ros.2015, Bera.2015}.
Deep in the localized phase, the $\ell$-bits have large overlaps with the physical spins.
Consequently, in a global quench from a local product state, such as the Néel state, the antiferromagnetic order remains nonzero at all times~\cite{Schreiber.2015}, as the overlaps of the $\ell$-bits with the initial state are constants of the motion~\cite{Serbyn.2013}.
In contrast, in thermalizing systems the antiferromagnetic order decays to zero due to the diffusion of magnetization (equivalently, transport of particles)~\cite{DAlessio.2016}.
Nonetheless, in localized systems the entanglement entropy grows slowly toward an extensive subthermal volume law~\cite{Znidaric.2008,Bardarson.2012}. 
This behavior has been explained by quasilocal (exponentially decaying) interactions between the $\ell$-bits, which induce entanglement through dephasing between different energy states~\cite{Serbyn.2013e8f}.
In this picture, there is no transport of particles, apart from an initial expansion of local densities within the single-particle localization length. 
This is consistent with early numerical work on many-body localization, which explored half-chain particle number fluctuations and found it to not increase with system size~\cite{Singh.2016}.

More recent work~\cite{Kiefer-Emmanouilidis2020-2} investigating particle transport via the number entropy $\SN$---the Shannon entropy of the particle number distribution in half of the chain---has found evidence for it increasing extremely slowly with time and saturating in finite systems at a value that increases with system size, albeit in a subextensive way.
The authors of this work interpreted their findings to imply the presence of transport and the absence of many-body localization.
While these first results were likely obtained either on the ergodic side of the transition~\cite{Luitz2020}, or in the purported prethermal finite-size crossover region~\cite{Morningstar.2022, Crowley.2022zf, Sels.2022, Long.2023}, later simulations deeper in the localized phase obtained consistent results~\cite{Kiefer-Emmanouilidis2021, Kiefer-Emmanouilidis2021.u, Kiefer-Emmanouilidis2022}. 
These observations have been suggested to arise from rare resonances and their associated transport~\cite{Ghosh2022}, though this was subject to debate~\cite{Kiefer-Emmanouilidis2022-comment-on-Ghosh2022,Ghosh2022-response-to-comment-on-Ghosh2022}.

Other recent work based on finite-size numerics has either cast doubt on the existence of many-body localization~\cite{Suntajs.2020g19,Suntajs.2020}, or reinterpreted the localized phase diagram to have a large nonlocalized crossover region, shifting the transition to the localized phase to a much larger disorder strength than hitherto believed~\cite{Morningstar.2022, Crowley.2022zf, Sels.2022, Long.2023}.
It can, however, be difficult to interpret finite-size numerics without a definite scaling theory of the localization transition~\cite{Abanin.2021, Sierant.2020}.
It also seems clear that, even if there is no many-body localization in a large region of disorder strengths, both finite-size numerics and finite-time experiments are dominated by the physics of many-body localization, albeit in some cases with an evident creep toward more extensive states~\cite{Doggen.20183z7, Weiner.2019zz2, Lezama.20199vg, Sierant.2022, Evers.2023}.
With more detailed observables and numerics, it seems pertinent to also refine our understanding of the many-body localization phenomenology.

Here we investigate to what extent the growth of number entropy can be explained directly within the phenomenology of many-body localization. 
Instead of working with microscopic models, we work directly with an effective $\ell$-bit model of localized interacting integrals of motion.
The $\ell$-bit model may be sufficiently rich in features to host phenomena beyond those initially understood.
In the $\ell$-bit model, particle number fluctuations arise both in the exponentially decaying support of the $\ell$-bits and from their exponentially decaying interactions.
While it is expected that particle transport in the localized phase is minimal, in line with earlier findings, it is not \textit{a priori} obvious that transport is entirely absent.
We therefore introduce a comprehensive $\ell$-bit model of many-body localization that takes into account both of these exponential tails, and measure the number entropy through numerical simulations following a quench from a local (in the physical spins) product state.
Since the Hamiltonian in the $\ell$-bit basis is diagonal, we can simulate to large times and for relatively large systems---computational limitations arise from the basis change from the $\ell$-bits to the physical spins, required for calculating observables.
Focusing on the quench dynamics, we show that the $\ell$-bit model exhibits ultraslow growth of the number entropy, which in a finite system saturates at a subextensive value that nevertheless increases with system size.
We find a growth in time consistent with $\ln \ln t$, though other functional forms cannot be ruled out.
We interpret this as resulting from the exponential tails in the $\ell$-bit model containing particle fluctuations.
Our observation of slow growth of number entropy implies that the $\ell$-bit model is richer than previously understood, and observation of such slow growth in microscopic models may therefore still be consistent with many-body localization.
%

\textit{Number entropy}---Our main observable of interest is the number entropy $\SN^A$, which is the contribution to the entanglement entropy $\SE^A$~\footnote{$\SE^A = -\mathrm{Tr}(\rho^A \log \rho^A)$ is the bipartite entanglement entropy where $A \cup \bar A$ is the whole system and $\rho^A = \mathrm{Tr}_{\bar A}(\rho^{A \cup \bar A}$) is the reduced density matrix. Equivalently, $\SE^A = -\sum_i (\lambda^A_i)^2 \log \left[ (\lambda^A_i)^2 \right]$ where the squared Schmidt values $(\lambda_i^A)^2 = (\lambda_i^{\bar A})^2$ equal the eigenvalues of $\rho^A$.} arising from the particle number fluctuation between subsystems $A$ and $\bar A$ (where $A\cup \bar A$ is the whole system)~\cite{Wiseman.2002, Melko.2016, Lukin.2019}.
We consider pure states $\ket{\psi}$ describing $L$ qubits labeled by $n_i \in \{0,1\}$ on each site $i$ of a one-dimensional lattice. For $A = [0\dots i]$, the number entropy is defined as $\SN^A = - \sum_n p_i(n) \ln p_i(n)$, where
\begin{equation}
	p_i(n) = \sum_{\substack{n_0\dots n_{L-1}, \\ \text{such that} \\ \sum_{k=0}^{i}n_k = n}} |\braket{n_0\dots n_{L-1}}{\psi}|^2,
    \label{eq:particle-number-probability}
\end{equation}
is the probability that $n$ qubits in $A$ are measured to be in the state $\ket{1}$.
From $p_i(n)$, we also obtain the probability $q_i(n)$ that the particle number grows from $n-1$ to $n$ at site $i$ (counting from the left):
\begin{equation}
    q_i(n) = \sum_{m=n}^{L} p_i(m) - p_{i-1}(m), \quad (n \geq 1),
\label{eq:particle-site-probability}
\end{equation}
where we set $p_{-1}(m)=0$. This satisfies $\sum_{i=0}^{L-1} q_i(n) = 1$ and $\sum_{n=1}^L q_i(n) = \langle n_i\rangle$, and can be used to calculate the moments of position $x$ for the $n$th particle, that is, $\langle x_n\rangle=\sum_{i=1}^L i q_i(n) $ and $\VarX=\sum_{i=1}^L (i - \langle x_n\rangle)^2  q_i(n)$~\cite{supplemental}.
We denote by $\SE$ and $\SN$ the half-chain entropies for the subsystem $A = [0\dots L/2)$.
Furthermore, we will refer to each $\ket{1}$ as particle, charge, or magnetic excitation interchangeably, and focus exclusively on systems where their number is conserved on $A\cup \bar A$.
Lastly, we interpret $\VarX>0$ as the spread of the $n$th particle $\ket{1}$ onto multiple sites by becoming dressed with particle-hole excitations, in which case we refer to it as a “quasiparticle”.
%


\textit{Model}---We consider Hamiltonians $H$ in the many-body-localized regime describing spin-$1/2$ chains of length $L$ with short-range interactions, in the sector of zero magnetization (half-filling).
This includes the disordered Heisenberg XXZ spin chain, which maps to the $t-V$ model of interacting fermions via the Jordan-Wigner transformation~\cite{Oganesyan.2007,Znidaric.2008,Bardarson.2012}.
In such Hamiltonians, \textit{physical spins} are the set of Pauli spin operators $\{\vec\sigma_i\}$ acting on the Hilbert space of site $i$.
A corresponding set of $\ell$-bits $\tau_i^z \equiv U \sigma_i^z U^\dagger$ emerges in the many-body localized phase, which are physical spins dressed by a quasilocal finite-depth unitary transformation $U$ that diagonalizes the Hamiltonian~\cite{Serbyn.2013,Huse.2014,Bauer.2013,Pekker.2017e}.

To study the growth of number entropy, we need to time evolve large spin chains to large times with high precision. 
We achieve this by time evolving matrix product states directly in the diagonal basis $H' = U^\dagger H U$, thereby avoiding Trotterization errors~\cite{Schollwoeck2010,supplemental},
\begin{align}
    \ket{\psi(t)} = U e^{-\mathrm{i}H't} U^{\dagger} \ket{\psi(0)}.
\end{align}
We thus model an \textit{effective} Hamiltonian $H$ by two random components: the diagonal $\ell$-bit Hamiltonian $H'$, which models the interactions between $\ell$-bits, and the transformation $U$ expressed as a finite-depth circuit of two-site gates.
The $\ell$-bit Hamiltonian is defined as~\cite{Huse.2014}
\begin{equation}
     H' = \sum_i h_i \tau_i^z  + \sum_{i < j} J_{ij} \tau_i^z \tau_j^z 
             + \sum_{i<j<k} J_{ijk} \tau_i^z \tau_j^z \tau_k^z + \cdots,
    \label{eq:l-bit-hamiltonian}
\end{equation}
where the $h$'s are random fields and $J$'s are random quasilocal couplings. On average, the couplings decay exponentially with their index diameter, such that $J_{ij} \propto e^{-|i - j|/\xiJtwo}$, $J_{ijk} \propto e^{-|i - k|/\xiJthree}$, and so on, where $\xiJtwo,\xiJthree \ll L$ are distinct characteristic length scales.
We implement $H'$ in the $\ell$-bit basis (where $\tau_i^z$ are diagonal Pauli matrices) keeping terms up to third order~\footnote{We have kept $J_{ijk}$ in the model, but our results are reproduced qualitatively by $h_i$ and $J_{ij}$.}:
\begin{align}
    h_i    & = R_i                        && i = 0,1,\dots, L-1,                 \label{eq:lbitH-hi}\\
    J_{ij} & = R_{ij} e^{-|i-j| / \xi_J}  && \text{when } 1 \leq |i-j| < L,      \label{eq:lbitH-Jij}\\
    J_{ijk}& = R_{ijk} e^{-|i-k| / \xi_J} && \text{when } |i-k| = 2,             \label{eq:lbitH-Jijk}
\end{align}
with $\xi_J = 1$ and $R$'s random variables drawn from a normal distribution $\mathcal N(0, \sigma_R^2 = 1)$.
We expect $\xi_J$ and the variance $\sigma_R^2$ to be related in a microscopic model, and find that setting them equal strikes a good balance between two detrimental extremes: setting $\xi_J < \sigma_R^2$ results in a steplike entropic growth because sites at distance $d$ and $d+1$ entangle at distinct timescales, whereas $\xi_J > \sigma_R^2$ leads to excessively overlapping timescales, making the growth process less clear and too short-lived~\cite{Znidaric.2018}.

\begin{figure}[tb]
    \raggedright
    \begin{minipage}{38.4mm}
        \subfloat{\includegraphics[keepaspectratio, height=44mm, valign=b]{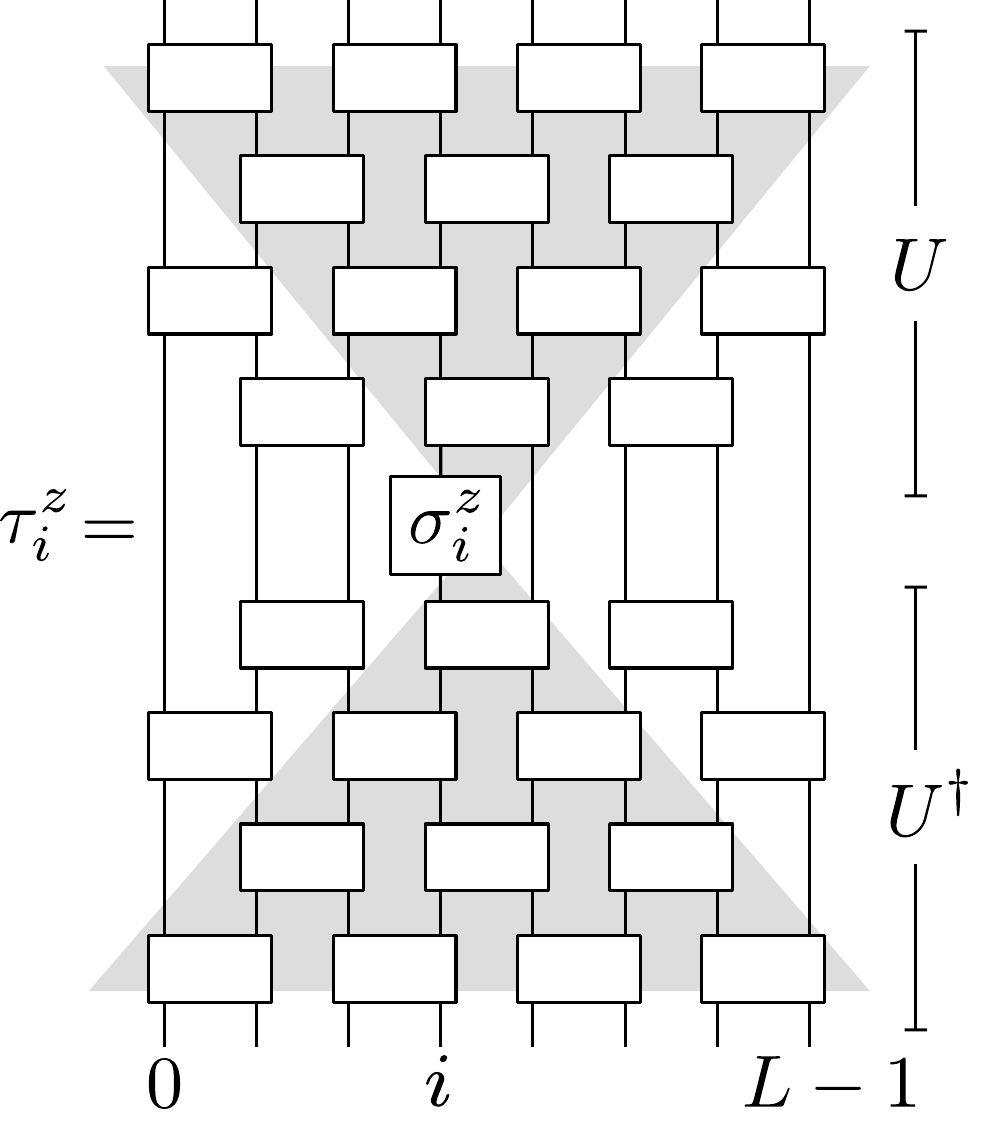}%
        \llap{\parbox[b]{73 mm}{\textbf{(a)}\\\rule{0ex}{38mm}}}%
        \label{fig:l-bit-unitary-circuit}}%
    \end{minipage}%
    \begin{minipage}{47.0mm}%
            \subfloat{\includegraphics[keepaspectratio,width=\linewidth,valign=b]{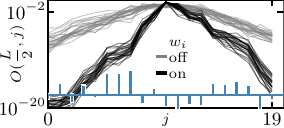}%
            \llap{\parbox[b]{7.0 mm}{\textbf{(b)}\\\rule{0ex}{15.30mm}}}%
            \label{fig:l-bit-support-weights}}\\
            \vspace{-3mm}
            \subfloat{\includegraphics[keepaspectratio,width=\linewidth,valign=b]{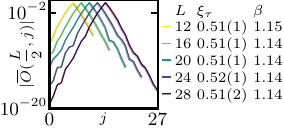}%
            \llap{\parbox[b]{48 mm}{\textbf{(c)}\\\rule{0ex}{15.30mm}}}%
            \label{fig:l-bit-support-decay}}%
    \end{minipage}\vspace{-2mm}%
    \caption{(a) An $\ell$-bit $\tau_i^z = U \sigma^z_i U^\dagger $ on $L=8$ sites with $U$ expressed as a “brickwork” circuit of two-site gates $u_i$ [white boxes, \cref{eq:unitary-gate}]. This circuit has depth $d_u=2$ because $U$ acts twice on each bond $i,i+1$. The range of the support (shaded gray) grows as $4d_u$.
    (b) The support of $\tau_{i=10}^z$ on $L=20$ sites for 25 circuit realizations with fixed $h_i$ (blue bars). This demonstrates the staggered decay caused by the weights $w_i$ [\cref{eq:circuit-weights}, black] where $|h_i-h_{i+1}|$ is large. With weights off ($w_i = 1$, gray), the decay is smooth.
    %
    (c) Supports for $\tau_{L/2}^z$ averaged over 10000 circuit realizations for each $L$, including the $w_i$ on and not fixing $h_i$. The parameters $\xi_\tau$ and $\beta$ have been extracted from stretched-exponential fits on all sites $i$. 
    In both (b) and (c) $f=0.2$ and $d_u=16$.
    }
    \label{fig:l-bits}
\end{figure}

To be consistent with the $\ell$-bit picture, and to draw direct parallels with studies of Heisenberg-like models, we require that $U$ is quasilocal and that it preserves the $U(1)$ symmetry.
The first condition means that the support of $\tau_i^z$ should decay exponentially in the physical basis.
The second implies that $U$ conserves the total magnetization (particle number).
These requirements are met if we express $U$ as a random unitary circuit with finite depth $d_u$, as depicted in \cref{fig:l-bit-unitary-circuit}, where each gate $u_i$ preserves the $U(1)$ symmetry.
We use $(L-1)\times d_u$ unique gates $u_i$ defined as unitary operators acting on sites $i$ and $i+1$
\begin{align}
    u_i ={}& e^{-\mathrm{i}f w_i M_i}, \label{eq:unitary-gate} \\
    w_{i} ={}& e^{-2|h_i - h_{i+1}|}, \label{eq:circuit-weights} \\
    \begin{split} \label{eq:hermitian-exponent}
        M_i ={}& \frac{\theta_1}{2} \sigma_i^z + \frac{\theta_2}{2} \sigma_{i+1}^z + \frac{\theta_3}{2} \lambda \sigma_i^z \sigma_{i+1}^z  \\
               & + c\sigma_i^+\sigma_{i+1}^- + c^*\sigma_i^-\sigma_{i+1}^+, 
    \end{split}
    %
\end{align}
where $f$ is a “mixing factor,” $w_i$ are weights discussed momentarily, and $M_i$ are random Hermitian matrices.
The mixing factor controls the exponential length scale $\xi_\tau$ with which $\tau_i^z$ decays (on average) away from site $i$.
We set $f=0.2$ and $d_u = 16$ and use these values unless stated otherwise.
This circuit depth allows $\ell$-bits to spread edge to edge up to $L=32$.
In \cref{eq:hermitian-exponent}, $\sigma^z$ and $\sigma^\pm = \sigma^x \pm i\sigma^y$ are $2\times 2$ Pauli matrices while $\theta_{1},\theta_2, \theta_3$, $\mathrm{Re}(c)$, and $\mathrm{Im}(c)$ are five normal random variables in $\mathcal N(0,1)$ drawn independently for each gate.
The first three $\theta$ terms along the diagonal mix the superposition of different configurations at fixed magnetization on sites $i,i+1$, while the last two off-diagonal $c$ terms exchange the magnetization of those two sites.
The parameter $\lambda$ controls the many-body content of $U$. 
All results in the main text use $\lambda = 1$.
The case $\lambda = 0$ corresponds to Anderson orbitals and would result in a model similar to the effective model introduced in Ref.~\onlinecite{Tomasi.2019}, though with different $\ell$-bit interaction structure. 
The number entropy in that model was studied in Ref.~\onlinecite{Kiefer-Emmanouilidis2022} where it was concluded that it failed to explain the growth of number entropy in microscopic models.
We find that the value of $\lambda$ does not qualitatively affect our results, even at $\lambda = 0$~\cite{supplemental}.

The weights $w_i$ in \cref{eq:unitary-gate,eq:circuit-weights} are introduced such that the spread of the $\ell$-bits is suppressed when the neighboring fields $h_i$ from \cref{eq:lbitH-hi} differ substantially.
The effect of $w_i$ is seen in \cref{fig:l-bit-support-weights}, where we have used the operator overlap $O(i,j) = \mathrm{Tr}(\tau_i^z \sigma_j^z)/2^L$ to measure the support of $\ell$-bits in the physical basis~\cite{supplemental,Chandran2015, Rademaker2017}.
Upon disorder-averaging, \cref{fig:l-bit-support-decay} shows that weights reproduce the expected decay from the $\ell$-bit picture, that is, an exponential ($\beta \approx 1$) with a characteristic length scale $\xi_\tau \approx 0.5$, comparable to previous estimations deep in the many-body localized phase~\cite{Kulshreshtha.2018}. 
In contrast, with the weights disabled ($w_i=1$), we typically get a larger stretching exponent $\beta$~\cite{supplemental}.
The weights can be heuristically motivated by variable-range hopping in the context of Anderson localization~\cite{Mott.1968,Banerjee.2016,Girvin-Yang.2019}.
In this conduction mechanism, the transition amplitude $\mathcal T$ between two states localized at distance $d$ with energies $E_a$ and $E_b$ is said to decay exponentially as $\mathcal T \sim  e^{-d/\xi} e^{-|E_a-E_b|/{k_B T}}$, where $\xi$ is the characteristic length scale of localization. 
In our case, the circuit structure models the first exponential.
To model the second exponential, we use the fields $h_i$ as a proxy for energies, since these are dominant in $H'$ and correspond to the single-particle energies in the noninteracting limit. 
The use of weights results in the distributions of saturation values of $\SE$ and $\SN$ becoming qualitatively similar to that of the disordered Heisenberg XXZ model~\cite{Luitz2020}.
We verified that our main conclusions, in particular on the growth of number entropy, do not depend on these weights.
We note that the quasilocality of $\ell$-bits manifests in two characteristic length scales $\xi_J$ and $\xi_\tau$ that are determined independently in $H'$ and $U$, respectively.
In microscopic models, these two length scales are typically correlated and much smaller than $L$ in the many-body localized regime.
%

\begin{figure*}
    \raggedright
    \setlength{\fboxsep}{0pt}%
    \captionsetup[subfigure]{margin=0pt}%
    \subfloat{%
            \includegraphics[width=59.5mm]{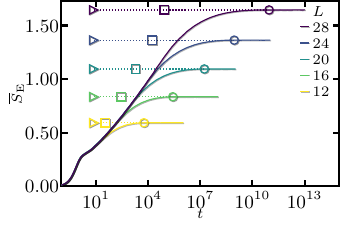}%
            \label{fig:SE-lnt}%
            \begin{picture}(0,0)%
                \put(-69,19.5){\includegraphics[trim=1.5mm 0mm 0mm 0mm,clip,height=15mm]{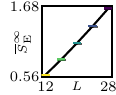}}%
            \end{picture}%
    }%
    \llap{\parbox[b]{102mm}{\textbf{(a)}\\\rule{0ex}{38mm}}}%
    \subfloat{%
            \includegraphics[width=59.5mm]{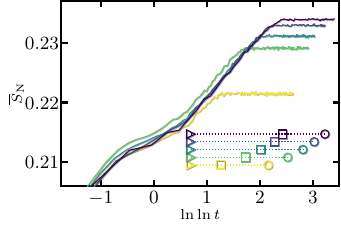}%
            \label{fig:SN-lnlnt}%
            \begin{picture}(0,0)%
                \put(-135,60.90){\includegraphics[height=17.5mm]{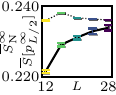}}%
            \end{picture}%
    }%
    \llap{\parbox[b]{102mm}{\textbf{(b)}\\\rule{0ex}{38mm}}}%
    \subfloat{%
            \includegraphics[width=59.5mm]{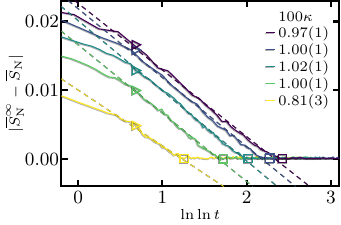}%
            \label{fig:SN-appr}%
    }%
    \llap{\parbox[b]{102mm}{\textbf{(c)}\\\rule{0ex}{38mm}}}%
    \hfill%

    \vspace{-3mm}%
    \subfloat{%
            \includegraphics[width=59.5mm]{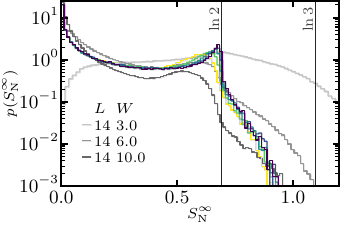}%
            \llap{\parbox[b]{7.8cm}{Ref.~\onlinecite{Luitz2020}:\\\rule{0ex}{22mm}}}%
            \label{fig:SN-dist}%
    }%
    \llap{\parbox[b]{102mm}{\textbf{(d)}\\\rule{0ex}{38mm}}}%
    \subfloat{%
            \includegraphics[width=59.5mm]{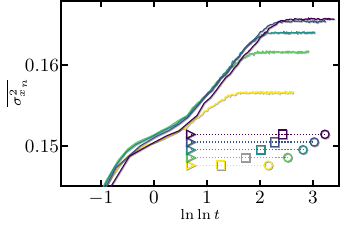}%
            \label{fig:VarX-lnlnt}%
            \begin{picture}(0,0)%
                \put(-135,60.90){\includegraphics[height=17.5mm]{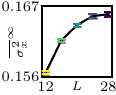}}%
            \end{picture}%
    }%
    \llap{\parbox[b]{102mm}{\textbf{(e)}\\\rule{0ex}{38mm}}}%
    \subfloat{%
            \includegraphics[width=59.5mm]{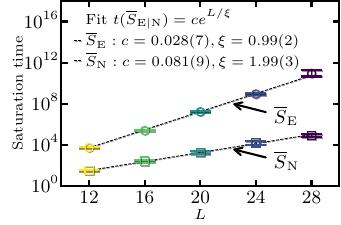}%
            \label{fig:SESN-tsat}%
    }%
    \llap{\parbox[b]{102mm}{\textbf{(f)}\\\rule{0ex}{38mm}}}%
    \hfill%
    \vspace{-3mm}
    \caption{We time evolve the Néel initial state (in the physical basis) using our $\ell$-bit model for system sizes $L=12$ up to $28$. 
    The legend in (a) applies in all panels.
    The half-chain entropies $\SE,\SN$, and variance $\VarX~$ are measured at $N_t=200$ logarithmically spaced time points $t$ for $L \leq 20$, and $N_t=100$ otherwise, spanning $0.1 \leq t \lesssim e^{L/\xi_J}$ (rounded up to a power of ten).
    Disorder averages $\davgSE,\davgSN$, and $\davgVarX$ are from $2\cdot 10^5$ random realizations of $H',U$ pairs for $L\leq 20$ and $10^5$ for $L > 20$.
    Saturation values are denoted $\SE^\infty$,$\SN^\infty$ and $\VarXInfty$. 
    Symbols indicate when $\davgSE\sim\ln t$ and $\davgSN \sim \ln\ln t$ both begin ($\whtrhd$), and when $\davgSN$ ($\whtsqr$) and $\davgSE$ ($\whtcrc$) saturate.
    (a,b,e) Insets show saturation values as functions of $L$ with bars denoting the standard error of the disorder average.
    (c) The approach to saturation values are fitted to $y = a - \kappa \ln \ln t$ (dashed) within the marked time intervals.
    (d) The distributions $p(\SN^\infty)$ are compared to results from Ref.~\onlinecite{Luitz2020} (gray) at fixed $L=14$ and various disorder strengths $W$. We use 100 bins in $0 \leq \SN^\infty \leq 1.2$ for the histogram.
    (e) The disorder-averaged variance of position $\davgVarX$ averaged for two central quasiparticles $n=L/4-1$ and $L/4$.
    (f) The saturation times $t_\whtsqr(\davgSE)$ and $t_\whtcrc(\davgSN)$---defined as the time when the running average $S_k = (N_t-k)^{-1}\sum_k^{N_t} \overline S_\mathrm{E|N}(t_k)$ ceases to grow monotonically---are estimated from 500 bootstrap samples of the disorder realizations.
    }
    \label{fig:results}
\end{figure*}

\textit{Numerical results}---In \cref{fig:results}, we present numerical results from simulated quenches using our model of interacting $\ell$-bits for system sizes up to $L=28$ and for at least $10^5$ random realizations of $H'$ and $U$.
Every simulation is initialized in the physical basis in one of the two antiferromagnetic Néel states ($\ket{0101\cdots}$ or $\ket{1010\cdots}$) chosen randomly. 
Observables are then measured during time evolution at logarithmically spaced time points spanning the entire process of entropic growth, until saturating due to finite system size.
The infinite-time saturation values, denoted with a superscript $\infty$, are estimated by time-averaging the corresponding time series after it has fully saturated.
Symbols $\whtrhd \csqr \whtsqr \csqr \whtcrc$ indicate the timescales for entropic growth and subsequent saturation (see the caption), as discussed further below. 

\cref{fig:SE-lnt} shows that our approach recovers the slow growth of disorder-averaged entanglement entropy consistent with $\davgSE \sim \xi_J \ln t$, expected in the many-body localized regime~\cite{Znidaric.2008,Bardarson.2012}.
The entropy saturates fully when $t \approx t_{\whtcrc}$ at values $\davgSE^\infty \sim L$ (inset), indicating a subthermal volume-law growth~\cite{Bardarson.2012,Nanduri.2014}. 

Our main result, presented in \cref{fig:SN-lnlnt}, demonstrates ultra\-slow growth of disorder-averaged number entropy consistent with $\davgSN \sim \ln \ln t$.
The inset shows that the saturation values $\davgSN^\infty$ (full lines) grow subextensively with $L$, though the functional dependence on $L$ cannot be extracted reliably from our data.
The Néel initial state was selected specifically to avoid spurious nonmonotonic growth of $\davgSN^\infty$ with $L$, observed with random initial product states~\cite{Kiefer-Emmanouilidis2021,Ghosh2022}. 
With random initial states our model shows the same nonmonotonicity, as we elaborate on in the Supplemental Material~\cite{supplemental}. 
The inset also shows the disorder average of $S[p_{L/2}^\infty(n)] = -\sum_n p_{L/2}^\infty(n) \ln p_{L/2}^\infty(n)$ (dotted line), which is the Shannon entropy of the saturation values of the particle number probabilities at the half-chain, $i=L/2$~\cite{Luitz2020}.
Since the entropy is a concave function of $p_i(n)$, we know from Jensen's inequality that $\overline{S}[p_{L/2}^\infty(n)]$ bounds $\davgSN$ from above~\footnote{
If $\phi(X)$ is a concave function of the random variable $X$, then Jensen's inequality $\langle \phi(X) \rangle \leq \phi(\langle X\rangle)$ holds. 
In this case we take $\langle \cdot \rangle$ to be the infinite-time average, that is, $\SN^\infty = \langle \SN(t > t_\whtsqr)\rangle$ on the left, and $S[p_{L/2}^{\infty}(n)] = S[\langle p_{L/2}(n, t > t_\whtsqr)\rangle]$ on the right of the inequality, which holds for every disorder realization.}.
Following an initial rise when $L \leq 16$, we find that $\overline S[p_{L/2}^\infty(n)]$ then decreases slowly until $L=28$, suggesting that $\davgSN$ could eventually saturate with system size.
However, as for $\davgSN$, we cannot extract a functional dependence on $L$ for $\overline S[p_{L/2}^\infty(n)]$. 
Since both $\davgSN$ and $\overline{S}[p_{L/2}^\infty(n)]$ are small averages from long-tailed distributions, these may require more disorder realizations to resolve the $L$ dependence precisely.

In \cref{fig:SN-appr}, the approach of $\davgSN$ to its saturation value $\davgSN^\infty$ strongly suggests that this growth is consistent with $\davgSN \sim \ln\ln t$. 
The linear fits to $y = a - \kappa \ln \ln t$ (dashed lines) agree convincingly within the time interval $t_\whtrhd < t < t_\whtsqr$ with slope $\kappa \approx 10^{-2}$. 
In contrast, a power law fit $\sim t^{-\alpha}$, argued for in Ref.~\onlinecite{Ghosh2022}, only agrees for shorter time intervals~\cite{supplemental}.

In \cref{fig:SN-dist} we compare the distribution of $\SN^\infty$ with results from Ref.~\onlinecite{Luitz2020} (in gray) that studied the disordered isotropic Heisenberg model.
Our results are in qualitative agreement with their high-disorder ($W\geq 6$) data.
The use of weights in \cref{eq:circuit-weights} has been crucial for attaining this agreement~\cite{supplemental}.
We recover the finite $L$ drift of a peak towards $\ln 2$, though our peaks increase slowly while becoming narrower as $L$ increases. 
The sharp drop before $\SN = \ln 3$ has been observed previously, and has been interpreted as a sign that $\SN$ stems mainly from the oscillations of a single particle~\cite{Luitz2020,Kiefer-Emmanouilidis2021}.

In \cref{fig:VarX-lnlnt} we plot the disorder-averaged width of quasiparticles at the center of the chain over time, where the width is estimated with the variance of position $\VarX$---see \cref{eq:particle-site-probability} and the surrounding discussion.
We take the average width of \textit{two} central quasiparticles, numbered $n=L/4-1$ and $n=L/4$, to account for the asymmetry of the particle positions inherent in the Néel states.
The growth is consistent with $\davgVarX\sim \ln \ln t$ on a time interval that is commensurate with the growth of $\davgSN$.
This time dependence is further supported by the approach to saturation, which admits a linear fit similarly to \cref{fig:SN-appr}, albeit with a smaller slope $\kappa \approx 0.0075$~\cite{supplemental}. 
The inset shows that saturation values $\davgVarXInfty$ grow monotonically but subextensively with $L$, similarly to $\davgSN^\infty$.
The case is markedly different for the position expectation values $\langle x_n\rangle$, which slowly drift back to the initial position following a quick shift in the early ballistic time $t < t_\whtrhd$~\cite{supplemental}.

Finally, we address the timescales involved during the growth and saturation of entropies. 
In the time interval where $\davgSE\sim\ln t$ holds, increasingly distant physical spins become entangled as they interact through the exponentially decaying couplings in $H'$, giving rise to the logarithmic entanglement light cone~\cite{Kim.2014,Deng.2017}.
The time interval is bracketed at $t < t_\whtrhd \approx 1$ by ballistic spreading of entanglement within the localization length  (approximately the inverse nearest-neighbor couplings) such that $\davgSE\sim t$, and much later, by the onset of saturation due to finite size.
Although we cannot determine precisely the time point when $\davgSE$ begins to saturate, \cref{fig:SE-lnt} suggests that it coincides with $t \approx t_\whtsqr$, the time when $\davgSN$ has saturated completely.
From \cref{fig:SESN-tsat}, the time $t_\whtsqr \sim e^{L/2}$ is comparable to $t \sim J_{ij}^{-1}$ at distance $|i-j| = L/2$ [see \cref{eq:lbitH-Jij}]. 
From the light cone perspective, this is the timescale for entanglement to spread over a distance $L/2$, from the half-chain to the edge, after which finite-size effects become noticeable.
\cref{fig:SE-lnt,fig:SESN-tsat} show that the saturation time for $\davgSE$ is consistent with $t_\whtcrc \sim e^{L}$ which, following the same argument, is the timescale for entanglement to spread edge to edge.

In \cref{fig:SN-lnlnt,fig:SN-appr}, we see that the ultraslow growth $\davgSN\sim \ln \ln t$ occurs in the time interval $t_\whtrhd < t < t_\whtsqr$.
Similarly to $\davgSE$, we again find ballistic growth $\davgSN \sim t$ in the initial time period $t < t_\whtlhd \approx 1$.
This is not a coincidence, since $\SN$ constitutes the largest contribution to $\SE$ during this time.
In contrast to the lengthy saturation process of $\davgSE$, the ultraslow growth of $\davgSN$ saturates abruptly when $t \approx t_\whtsqr$.

We see in \cref{fig:VarX-lnlnt} that the time points $t_\whtrhd$ and $t_\whtsqr$ apply to the quasiparticle width $\davgVarX$ as well.
The initial growth of both entropies can therefore be understood from the perspective of a quasiparticle next to the half-chain boundary: starting entirely localized on one site, the quasiparticle spreads ballistically within the localization length such that a small fraction crosses the boundary, thus generating a proportional amount of number entropy~\cite{supplemental}.
We observe the simultaneous saturation of $\davgVarX$ and $\davgSN$ at time $t_\whtsqr \sim e^{L/2}$, which we take as yet another indication that the ultraslow growth of both quantities are directly related to each other.  
%

\textit{Discussion}---We have demonstrated that a circuit model of interacting $\ell$-bits exhibits ultraslow growth in time of the half-chain number entropy. 
The observed growth is consistent with $\davgSN \sim \ln\ln t$ within a time interval that is exponential in system size.
Additionally, we have observed the monotonic and subextensive growth of saturation values $\davgSN^\infty$ with system size up to $L=28$.
However, due to the limited range of available system sizes, the functional form of $\davgSN^\infty(L)$ remains elusive.
The observation that the entropy $\overline S[p_{L/2}^\infty(n)]$, which bounds the number entropy from above, decreases in the range $L \in [16,28 ]$ indicates the possibility that $\davgSN^\infty$ may eventually saturate with $L$.

We have studied the quench dynamics from the perspective of quasiparticles---particles dressed by particle-hole excitations---that are characterized by their position $\overline{\langle x_n \rangle}$ and variance $\davgVarX$.
We discovered a robust connection between the behaviors of number entropy and quasiparticle variance.
Specifically, quasiparticles widen ultraslowly, with $\davgVarX \sim \ln \ln t$ (for half-chain particles) simultaneously with $\davgSN \sim \ln \ln t$, while $\overline{\langle x_n \rangle}$ remains near the initial position~\cite{supplemental}.
This suggests that the main contribution to $\SN$ comes from a single quasiparticle oscillating across the half-chain boundary, and crucially, that the growth of $\davgSN$ is caused primarily by the ultraslow widening of the central quasiparticle.
Notably, the applicability of the quasiparticle picture is exclusive to many-body localized systems, as generic states of ergodic systems are extended.
This suggests the possibility for distinct behaviors of $\davgSN$ vs.\ $L$ in localized and ergodic systems, irrespective of the observed ultraslow temporal growth in both cases.

As our model is many-body localized by construction, the ultraslow growth of number entropy is inherent to many-body localized systems, at least for the range of system sizes available to us.
An observation of ultraslow growth of number entropy in microscopic models is therefore not, by itself, sufficient evidence that such systems must delocalize in the thermodynamic limit. 
It remains a possibility that the thermodynamic limit will differ between our $\ell$-bit model and microscopic models; resolving this requires further (currently unattainable) studies.
Our random circuit $\ell$-bit model opens the door for more refined examinations of the phenomenology within the many-body localized phase.
Future studies should explore whether such comprehensive models can capture other phenomena that are traditionally attributed to the absence of many-body localization.
%

\textit{Acknowledgements}---We thank M.~Kiefer and J.~Sirker for helpful discussions.
This work received funding from the European Research Council (ERC) under the European Union’s Horizon 2020 research and innovation program (Grant Agreement No.~101001902). Thomas Klein Kvorning’s research is funded by the Wenner-Gren Foundations. Loïc Herviou has been supported by the Swiss National Science Foundation (FM, ZJ) grant 212082.  The computations were enabled by resources provided by the National Academic Infrastructure for Supercomputing in Sweden (NAISS) and the Swedish National Infrastructure for Computing (SNIC) at Tetralith partially funded by the Swedish Research Council through grant agreements no.~2022-06725 and no.~2018-05973.

\nocite{Chandran2015}
\nocite{Rademaker2017}
\nocite{Kulshreshtha.2018}
\nocite{Verstraete.2006}
\nocite{Bauer.2013}
\nocite{Friesdorf.2015}
\nocite{Vidal2003}
\nocite{Vidal2004}
\nocite{Schollwoeck2010}
\nocite{Stoudenmire2010}
\nocite{Kiefer-Emmanouilidis2021}
\nocite{Ghosh2022}
\nocite{Luitz2020}
\nocite{Kiefer-Emmanouilidis2022}
\nocite{Bardarson.2012}
\nocite{Bera.2015}
\nocite{Bera.2017}

\bibliography{references.bib}


\end{document}


\title{Supplemental Material for \texorpdfstring{\\}{}
Ultraslow Growth of Number Entropy in an \texorpdfstring{$\ell$-bit}{l-bit} Model of Many-Body Localization}
\author{David Aceituno Chávez}
\affiliation{Department of Physics, KTH Royal Institute of Technology, Stockholm, 106 91 Sweden}
\author{Claudia Artiaco}
\affiliation{Department of Physics, KTH Royal Institute of Technology, Stockholm, 106 91 Sweden}
\author{Thomas Klein Kvorning}
\affiliation{Department of Physics, KTH Royal Institute of Technology, Stockholm, 106 91 Sweden}
\author{Loïc Herviou}
\affiliation{Univ. Grenoble Alpes, CNRS, LPMMC, 38000 Grenoble, France}
\affiliation{Institute of Physics, Ecole Polytechnique Fédérale de Lausanne (EPFL), CH-1015 Lausanne, Switzerland}
\author{Jens H. Bardarson}
\affiliation{Department of Physics, KTH Royal Institute of Technology, Stockholm, 106 91 Sweden}

\maketitle

\appendixpageoff
\appendixtitleoff
\setcounter{figure}{0}    
\setcounter{section}{0}    
\makeatletter
\renewcommand{\appendixtocname}{Supplementary material}
\renewcommand\thesection{S\@arabic\c@section}
\renewcommand\thetable{S\@arabic\c@table}
\renewcommand\thefigure{S\@arabic\c@figure}
\renewcommand\theequation{S\@arabic\c@equation}
\makeatother



\section{Quasilocality of \texorpdfstring{$\ell$-bits}{l-bits}}
\label{sec:quasilocality}
\begin{figure}[t]
    \centering
    \includegraphics[clip, trim=0cm 0.5cm 10.0cm 0cm,width=0.70\linewidth, valign=c]{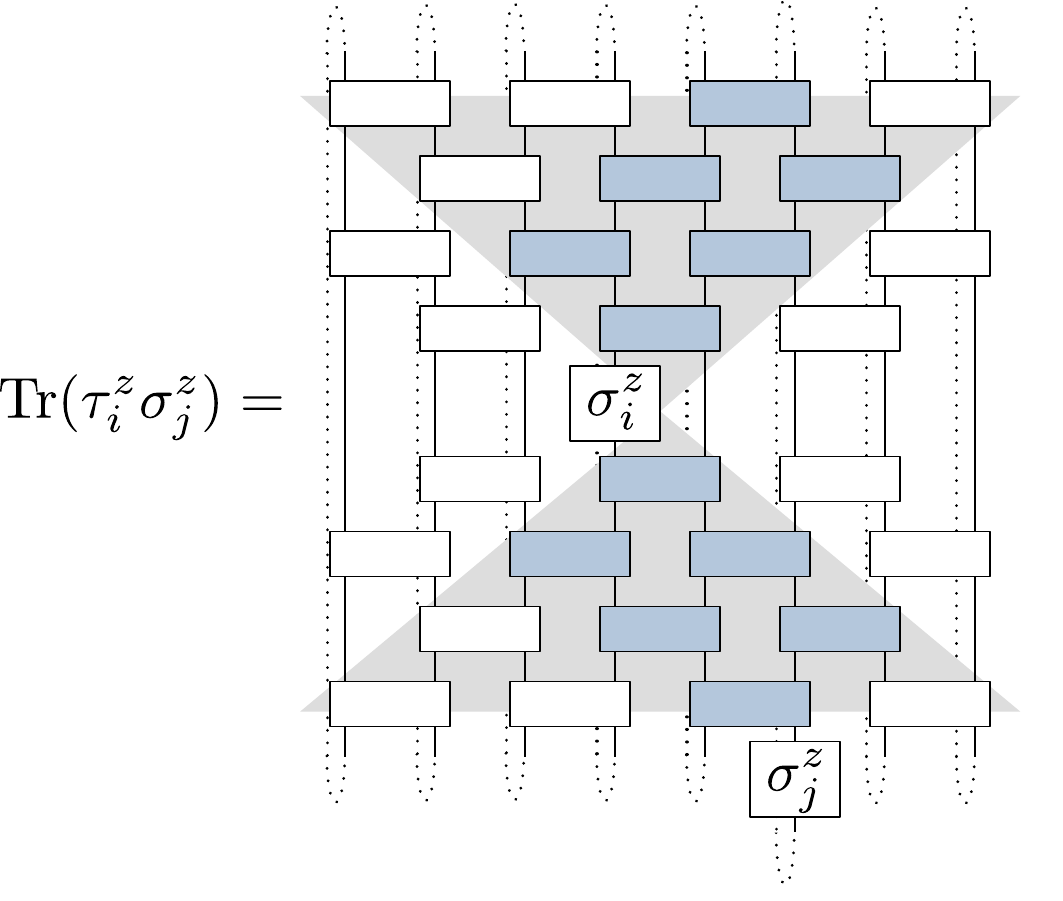}
    \vspace{-0.35cm}
    \caption{
    The operator overlap $O(i,j)$ [\cref{eq:operator-overlap}] measures the support of the $\ell$-bit $\tau_i^z$ at site $j$. The dotted lines show the tracing over free indices that takes place after contracting the circuit with $\sigma_j^z$. Only blue gates give a relevant contribution in this example because white gates contract to identities.
    }
    \label{fig:l-bit-overlap}
\end{figure}
\begin{figure}[t]
    \setlength{\fboxsep}{0pt}%
    \vspace{-0.35cm}
    \subfloat{
        \includegraphics[height=57mm]{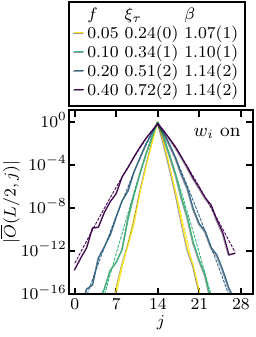}
        \llap{\parbox[b]{57mm}{\textbf{(a)}\\\rule{0ex}{34mm}}}
        \label{fig:l-bit-f-decay-arithmetic-won}
    }%
    \hspace{-0.55cm}
    \subfloat{
        \includegraphics[height=57mm,trim={11.0mm 0 0 0},clip ]{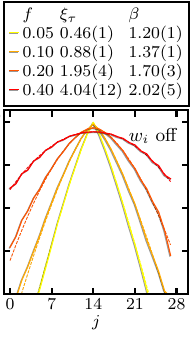}
        \llap{\parbox[b]{57mm}{\textbf{(b)}\\\rule{0ex}{34mm}}}
        \label{fig:l-bit-f-decay-arithmetic-woff}
    }
    \vspace{-0.35cm}
    \caption{The disorder-averaged support of $\tau_{L/2}^z$, calculated with \cref{eq:operator-overlap}, for system size $L=28$ and circuit depth $d_u=16$, while varying the mixing factor $f$.
    In (a) and (b) we compare the weights on ($w_i = e^{-2|h_i-h_{i+1}|}$, see main text Eq.~[9]) and off ($w_i = 1)$.
    %
    For each choice of parameters, we average 10000 realizations of the random unitary circuit and onsite fields $h_i$. 
    The dashed lines show the fit to a stretched exponential used to calculate $\xi_\tau$ and $\beta$. 
    }
    \label{fig:l-bit-decay}
\end{figure}

The quasilocality of the $\ell$-bit operators $\tau_i^z$ implies that their support on the physical basis should decay, on average, as $\sim e^{-|i-j|/\xi_\tau}$, where $\xi_\tau$ is the $\ell$-bit \textit{localization length}. 
Our circuit, which is defined in Eqs.~(8-10) in the main text, sets $\xi_\tau$ indirectly through the mixing factor $f$. 
We estimate $\xi_\tau$ by measuring the decay of $\tau_i^z$ in the physical basis using the operator overlap \cite{Chandran2015, Rademaker2017}
\begin{align}
    O(i,j) = \frac{\text{Tr} (\tau_i^z \sigma_j^z)}{2^L};
    \label{eq:operator-overlap}
\end{align}
the trace is computed using the circuit as depicted in \cref{fig:l-bit-overlap}. 
Next, we compute the arithmetic disorder average $\davg O(i,j)$ over many random realizations of the circuit as well as field configurations $h_i$.
Since the $\tau_i^z$ are localized, the matrix $\davg O(i,j)$ is peaked along the diagonal $i=j$.
By averaging the decay to the left and right of the diagonal entry we obtain $\davg O(i,|i-j|)$, and finally, by averaging over $i$, we compute $\davg O(|i-j|)$.
The localization length $\xi_\tau$ is obtained by fitting to a stretched exponential $|\davg O(x)| = C e^{-(x/\xi_\tau)^\beta}$, where $x=|i-j|$ and $\beta$ is the \textit{stretching exponent}. 
Examples of $|\davg O(i,j)|$ and fits are shown in \cref{fig:l-bit-decay} for a range of mixing factors $f$.

The calculation outlined above is used to guide the construction of the unitary circuit. 
In particular, we use it to ensure that $\beta$ is close to unity, since it would otherwise indicate a departure from the $\ell$-bit picture, where the operators should decay exponentially on average.
This is an important motivation for using the field-dependent weights $w_i$ introduced in Eq.~(9) in the main text.
Indeed, we see in \cref{fig:l-bit-f-decay-arithmetic-won} that $\beta$ remains close to unity for a broad range of $f$ when weights are enabled.
On the other hand, in \cref{fig:l-bit-f-decay-arithmetic-woff} we see that $\beta$ grows rapidly beyond unity with increasing $f$ when the weights are disabled (that is, $w_i=1$).
The decay of $|\davg O(x)|$ also guides the choice of $f$ because of its relation to $\xi_\tau$.
If $\xi_\tau$ is too small, we suppress the $\ell$-bit interactions whose dynamics we wish to investigate. 
On the other hand, if $\xi_\tau$ is too large, entanglement may proliferate to a point where it is no longer commensurate with the physics of localized systems, on top of escalating computational demands.
From a previous work, we also expect that $\xi_\tau \lesssim 1 \ll L$ deep in the many-body localized phase~\cite{Kulshreshtha.2018}.
Considering all these criteria, we find that $f = 0.2$ (with weights enabled) gives reasonable values for $\xi_\tau$ and $\beta$.

Finally, we can use $\davg O(|i-j|)$ to check that the tails of the $\ell$-bit supports are not overly truncated. 
This is only a concern in the case of shallow circuits, that is, with $d_u < L/2$, because deeper circuits can always accommodate the full diameter of $\ell$-bits on any site.
Even though a shallow circuit with carefully chosen parameters could improve the computational efficiency, we have opted to eliminate this source of error entirely by setting $d_u = 16$, which exceeds the longest half-chain considered here.
%


\section{Time Evolution}
\label{sec:time-evolution}
We are interested in studying the entire dynamical process following the quench of a disordered quantum system modeled by interacting $\ell$-bits.
Our approach is to time-evolve in the diagonal $\ell$-bit basis and perform measurements in the physical basis, transforming between these bases using a finite-depth random unitary circuit.
We use the matrix product states (MPS) ansatz to express quantum states at all times~\cite{Verstraete.2006,Bauer.2013,Friesdorf.2015}.

\subsection{Time step}
\label{sec:time-step}

Because the $\ell$-bit Hamiltonian $H'$ is diagonal (see Eq.~(4) in the main text), the time evolution can be performed in a single step without Trotterization and with an arbitrary time step $t$:
\begin{align}
    \ket{\psi'(t)} &= e^{-\mathrm{i}H't}\ket{\psi'(0)}.
    \label{eq:lbit-time-evolution}
\end{align}
More precisely, since all terms in $H'$ commute with each other, the exponential in \cref{eq:lbit-time-evolution} is trivially factorized into operators that act independently on the Hilbert space of the corresponding sites, that is,
\begin{align}
   \ket{\psi'(t)} &= \prod_i e^{-\mathrm{i}H_i't} \prod_{i<j} e^{-\mathrm{i}H_{ij}'t} \cdots \ket{\psi'(0)}.
   \label{eq:lbit-time-evolution-prod}
\end{align}
After applying the time-evolution operators, the state is brought back to an MPS form using singular value decomposition (SVD), as done, for example, in the time-evolving block decimation (TEBD) algorithm~\cite{Vidal2003,Vidal2004,Schollwoeck2010}. 
In fact, except for the lack of Trotterization and arbitrary time step, this part of our algorithm is identical to TEBD.

A single simulation calculates a series of states at increasing time points $\ket{\psi'(t_1)}, \ket{\psi'(t_2)},\dots \ket{\psi'(t_\text{max})}$.
Each time-evolved state $\ket{\psi'(t_n)}$ is obtained independently by applying the time evolution operator to the initial state $\ket{\psi'(0)}$ once. 
Following each time step, we use $U$ to transform the state back to the physical basis, where observables are measured.
Thus, a simulation can be summarized as
\begin{align}
    \ket{\psi(t_n)} &= U e^{-\mathrm{i}H't_n} U^{\dagger} \ket{\psi_\text{init}},
\end{align}
where $t_n$ are logarithmically spaced time points from $t_1=10^{-1}$ up to $t_{\max} \gg e^{L/\xi_J}$.
This choice of $t_{\max}$ exceeds the time scale $J_{ij}^{-1}$ of the smallest interactions in $H'$, at range $|i-j| = L-1$, after which the entropies should have fully saturated.

\subsection{Long-range terms}
\label{sec:swap-gates}
\begin{figure}[t]
    \centering
    \includegraphics[clip, trim=0cm 9.7cm 23.0cm 0cm,height=35mm, valign=c]{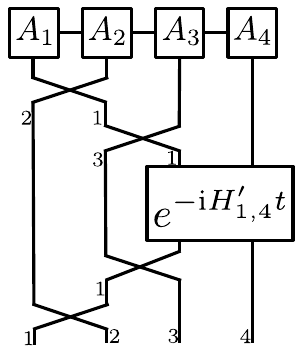}
    \caption{A sequence of swap-gates $s_{1,2}, s_{2,3}$, drawn as crossings, moves the state $A_1$ on site $1$ to site $3$. 
    The time-evolution operator can then act on a two-site MPS tensor $A_1\otimes A_4$ with spin dimension $d^2$, before swapping the spins back to their original sites. For long-range terms, this operation reduces the memory cost in exchange for swap operations.
    }
    \label{fig:swap-gates}
\end{figure}

We use the method of swap gates \cite{Stoudenmire2010} to apply time-evolution operators for long-range terms in $H'$.
In this method, a swap operation $s_{i,i+1}$ exchanges the physical indices of two neighboring sites of the MPS, such that indices $i$ and $i+1$ point to sites $i+1$ and $i$, respectively.
After exchanging indices, the combined matrix of sites $i$ and $i+1$ is split with an SVD, which dominates the runtime cost of this operation.
The core idea, depicted in \cref{fig:swap-gates}, is to sandwich each operator $e^{-\mathrm{i}H'_{ij}t}$ from \cref{eq:lbit-time-evolution-prod} between two sequences of swap gates.
The first sequence moves site $i$ towards $j$, so that $e^{-\mathrm{i}H'_{ij}t}$ can act on nearest neighbors (that is, on a subspace of size $2^2$ instead of $2^{j-i+1}$).
The second sequence moves site $i$ in reverse to its original position.
Note that, since repeated swaps cancel ($s_{i,i+1}^2 = 1$), it is possible to cancel sequences between successive long-range operators. 
Thus, by re-ordering the product in \cref{eq:lbit-time-evolution-prod}, one can cancel most swaps and reduce the computational overhead dramatically.
\subsection{Precision}
Our method avoids two sources of cumulative error found in conventional TEBD: the error due to the Trotter decomposition and the error from repeated truncation of singular values between incremental time steps.
Instead, states at different time points have the same upper bound on numerical error because they are fully independent of each other. 
This means that each time-evolved state is produced by a fixed number of truncations, namely after each site swap operator, after each time evolution operator [\cref{eq:lbit-time-evolution-prod}], and after each two-site gate in the unitary circuit (Eq.~(8) in the main text).

When truncating singular values, our strategy is to fix the truncation error threshold $\epsilon$, and let the bond dimension $\chi$ adapt freely to obtain constant precision at all times. 
More precisely, we require that the truncation error, defined as the norm of the discarded singular values $\lambda_\alpha$, be smaller than a fixed threshold $\epsilon$:
\begin{equation}
    \sqrt{\sum_{\alpha = \chi}^{r} \lambda_\alpha^2} <  \epsilon,
    \label{eq:truncation-error}
\end{equation}
where $r$ is the rank of the SVD and $\chi$ is chosen to satisfy the inequality on every individual truncation.
The value of $\epsilon$ translates to an observed error of the same magnitude in the entropies that we measure at each time point.
We have found in trials that $\epsilon = 10^{-5}$ is sufficiently small: smaller values increase computational demand without any discernible differences in our results.
Although $\chi$ saturates after some time due to finite system size, the level at which it saturates varies greatly for each disorder realization.
Therefore, the main disadvantage of fixing $\epsilon$ is that the memory usage for a given simulation is uncontrolled.
We have found empirically that the average bond dimension grows slowly enough with system size that the tradeoff is worthwhile for system sizes below $L\approx 32$, with present computational resources.
For the largest system sizes, a small fraction of realizations will outgrow their memory allocation, but these can simply be run again separately with more memory.

Lastly, we remark that since each term in $H'$ is diagonal, one can avoid costly matrix exponentials in \cref{eq:lbit-time-evolution-prod}.
Instead, one can take the exponential of each diagonal element $x$ directly, as $e^{-\mathrm{i}(xt\mod 2\pi)}$, where the modulo prevents numerical overflow in the exponential. 
We caution that the modulo yields fewer correct digits as $2\pi/xt$ gets closer to floating-point precision.
For example, $10^{7} \mod 2 \pi$ gives 1 or 9 correct digits, respectively, when using 32-bit or 64-bit IEEE 754 floating-point numbers which, in turn, fit approximately 7 or 16 decimal digits of precision, respectively. 
We use 128-bit “quadruple” precision for all the numbers involved in generating the time evolution operator. 
Since 128-bit precision fits 34 decimal digits, the error from the modulo is negligible in the largest time scale that we target here.

\section{Initial state}
\label{sec:initial-state}
\begin{figure}[tb]
    \centering
        \includegraphics[clip, trim=0cm 1.8cm 0cm 2.4cm,]{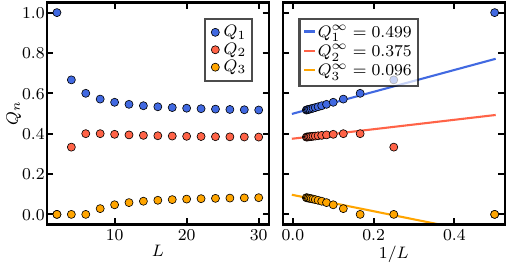}%
        \llap{\parbox[b]{146mm}{\textbf{(a)}\\\rule{0ex}{40mm}}}%
        \llap{\parbox[b]{14mm}{\textbf{(b)}\\\rule{0ex}{40mm}}}%
    \vspace{-0.3cm}
    \caption{(a) We compute the probability directly with $Q_n=N_n/N$, where $N$ is the total number of random product states of size $L$ (with zero magnetization) and $N_n$ is the subset of states having jump distance $n$ before contributing to $\SN$, as described in the text.
    (b) We extrapolate the tendency for $L > 10$ to obtain $Q_n^\infty = Q_n(L\rightarrow \infty)$.}
    \label{fig:q-distribution}
    \vspace{-0.2cm}
\end{figure}
We concentrate on initial states $\ket{\psi_\text{init}}$ with an extensive number of spins available for magnetic exchange, in the sector of zero total magnetization  such that $\sum_{i=1}^L \langle \sigma_i^z \rangle = 0$. 
This includes random product states (with zero magnetization) and Néel states (when $L$ is even), but not, for instance, states with a single domain-wall in the center.
While we observe $\davgSN\sim \ln\ln t$ with both random product states and Néel states, we find with the former that the saturation values $\davgSN^\infty$ grow nonmonotonically with $L$.
This nonmonotonicity has been noticed previously~\cite{Kiefer-Emmanouilidis2021,Ghosh2022}; in Ref.~\onlinecite{Ghosh2022} it has been attributed to the fact that an increase in $L$ is accompanied by a change in the probability $Q_n$ that, for a given state, the shortest distance between any particle and a hole on the opposite half-chain will be $n$.
The key idea is that states with shorter distance $n$ will have larger $\SN$, on average.
We depict $Q_n$ as a function of $L$ in \cref{fig:q-distribution}.
While in the limit $L\rightarrow \infty$ we expect $Q_1 = 0.5$, the probability is larger and varies significantly in the range of system sizes that is accessible to our numerical analysis. 
Given that $\davgSN^\infty$ is determined largely by how the central quasiparticle crosses the half-chain boundary, we have chosen the Néel initial state (which has $Q_1 = 1$ for all $L$) to eliminate the spurious nonmonotonicity in saturation values.

\section{Relation between \texorpdfstring{$\SN$ and $\sigma^2_{x_{n}}$}{}}
\label{sec:relation}
\begin{figure}
    \centering
    \includegraphics[width=\linewidth]{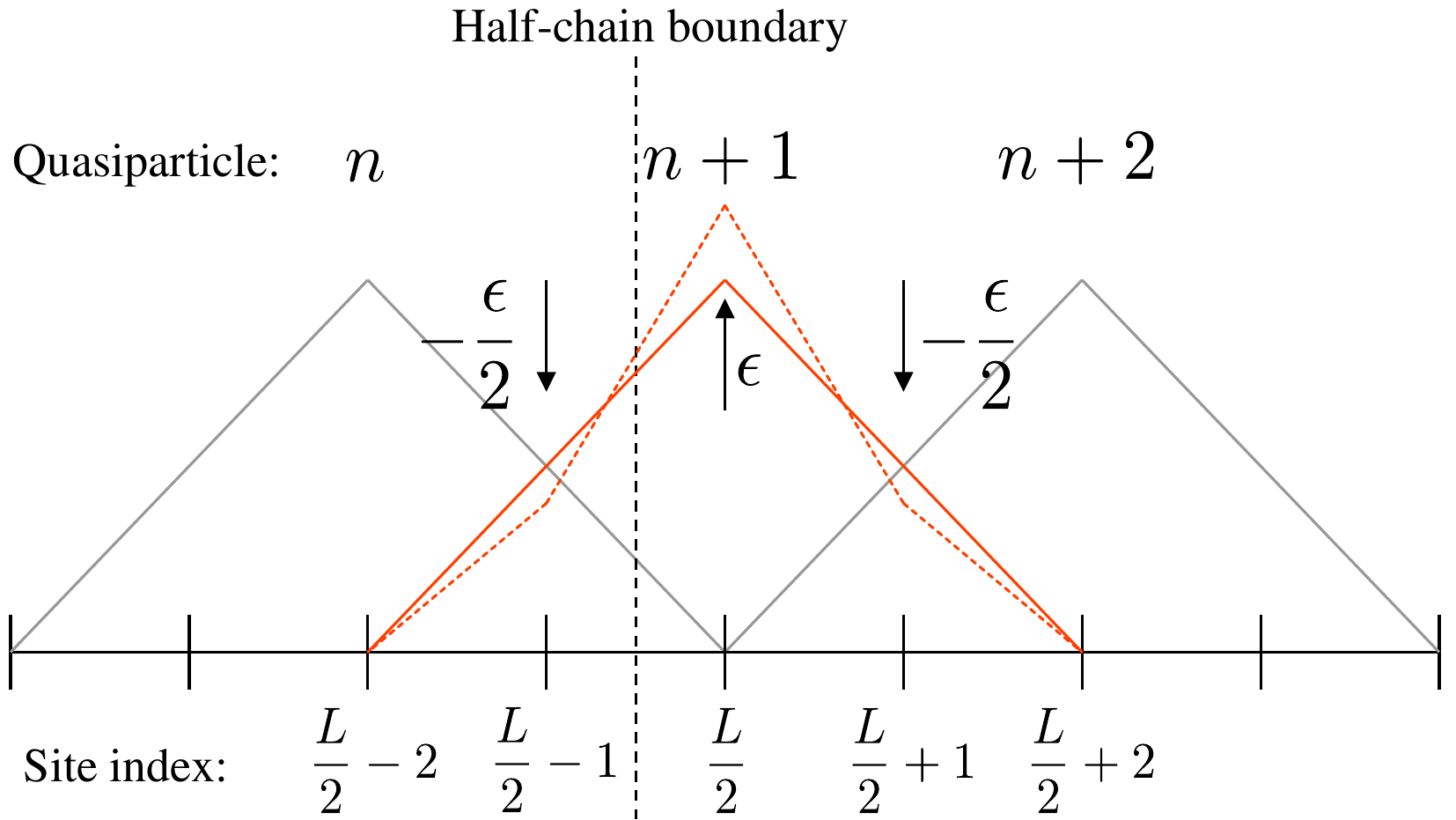}
    \caption{Perturbation of $q_i(n+1)$ on a Néel-like state.}
    \label{fig:perturbation}
\end{figure}

In the following, we establish a relationship between the number entropy $\SN$ and the quasiparticle width $\VarX$ that was introduced in the discussion following Eq.~(2) in the main text.
We show that a perturbation that only changes the width of the central quasiparticle, such that its variance $\sigma^2(x)$ changes by $-\epsilon$, will change $\SN$ by an amount $-\epsilon k + O(\epsilon^2)$, where $k$ is a positive proportionality factor that depends on the number probabilities $p_i(n)$ before the perturbation.

Consider the following perturbation of the central quasiparticle $n+1$
\begin{align}
    \Delta q_{L/2-1}(n+1)               &= -\epsilon/2 \label{eq:delta-q1}\\
    \Delta q_{L/2\hphantom{+0}}(n+1)    &= \epsilon \label{eq:delta-q2}\\
    \Delta q_{L/2+1}(n+1)               &=  -\epsilon/2,\label{eq:delta-q3}
\end{align}
where $\Delta q = \tilde q - q$ is the difference between the perturbed and unperturbed values. 
This perturbation reduces the width of the particle without changing the expectation value of its position $\langle x \rangle$, as shown in \cref{fig:perturbation}. Note that $\tilde q_i(n+1)$ remains normalized, since the changes cancel each other.
The change in the expectation value for the position of this quasiparticle is
\begin{align}
    \langle \tilde x \rangle &- \langle x \rangle = \nonumber
     \sum_{i=0}^{L-1} i \Delta q_i(n+1) \\
    &= (\frac{L}{2}-1)(-\frac{\epsilon}{2}) + \frac{L}{2}\epsilon +  (\frac{L}{2}+1)(-\frac{\epsilon}{2}) = 0, \label{eq:delta-x}
\end{align}
as expected.
Given that $\langle \tilde x \rangle = \langle x \rangle$, the corresponding change in the variance is
\begin{align}
    \tilde \sigma^2(x) - \sigma^2(x) &= 
    \sum_{i=0}^{L-1} (i - \langle x\rangle)^2 \Delta q_i(n+1) \nonumber \\
    &= \sum_{i=0}^{L-1}(i^2  - 2i\langle x \rangle  +  \langle x \rangle^2) \Delta q_i(n+1),
    \label{eq:delta-sigma}
\end{align}
where the first term gives
\begin{align}
    \sum_{i=0}^{L-1} & i^2 \Delta q_i(n+1) \\
    &= (\frac{L}{2}-1)^2(-\frac{\epsilon}{2}) + (\frac{L}{2})^2\epsilon +  (\frac{L}{2}+1)^2(-\frac{\epsilon}{2})\\
    &= -\epsilon,
\end{align}
while [using \cref{eq:delta-x}] the second term gives
\begin{align}
  2  \langle x \rangle    \sum_{i=0}^{L-1} i\Delta q_i(n+1) = 2\langle x \rangle (\langle \tilde x \rangle - \langle x \rangle ) = 0.
\end{align}
Finally, the remaining term in \cref{eq:delta-sigma} gives
\begin{align}
    \langle x \rangle^2 \sum_{i=0}^{L-1} \Delta q_i(n+1) = 0,
\end{align}
where the sum vanishes because the perturbations in \cref{eq:delta-q1,eq:delta-q2,eq:delta-q3} cancel each other. 
Thus, the change in variance is
\begin{equation}
    \tilde \sigma^2(x) - \sigma^2(x) = - \epsilon,
\end{equation}
confirming that the width reduces by $-\epsilon$. 

\begin{table}[t]
\begin{center}
    \begin{tabular}{| l | c | c | c |}\hline
     $\hphantom{\Delta}\Delta p_i(m)$         & $m=n$          & $m=n+1$      & $m=n+2$\\\hline
     $i=L/2-1$              & $-\epsilon$    & $\epsilon/2$ & $\epsilon/2$ \\\hline
     $i=L/2\hphantom{-1}$   & $-2\epsilon$   & $\epsilon$   & $\epsilon$ \\\hline
     $i=L/2+1$              & $-3\epsilon/2$ & $\epsilon/2$ & $\epsilon$ \\\hline
\end{tabular}
\end{center}
\vspace{-0.4cm}
\caption{}
\label{tab:concrete}
\end{table}
Next, we calculate the change in the half-chain number entropy $\Delta \SN = \tilde{S}_\mathrm{N} - \SN$ from the corresponding change in number probabilities $\Delta p_i(m)$ for $m = n-1, n$ and $n+1$.
There are many possible configurations $\Delta p_i(m)$ that would satisfy the present perturbation for $q_i(n+1)$, while leaving $q_i(n)$ and $q_i(n+2)$ unchanged, and maintaining normalization such that $\sum_{m} p_i(m) = 1 \; \forall i$ and $\sum_{i} q_i(m) = 1 \; \forall m$. \cref{tab:concrete}  gives a concrete example, where the normalization is maintained by virtue of each row adding to zero. 
For the half-chain number entropy, we only need the row for $i = L/2$. For brevity, we introduce the shorthands $p^m = p_{L/2}(m)$, $\tilde p^m = \tilde p_{L/2}(m)$  and $\delta^m = \Delta p_{L/2}(m) = \tilde p^m - p^m$:
\begin{align}
    \Delta \SN &= \sum_{m} -\tilde p^m \ln (\tilde p^m ) +  p^m \ln (p^m) \\
               &= \sum_{m} -p^m \ln\left(1 + \frac{\delta^m}{p^m}\right) - \delta^m\ln(p^m+\delta^m) ,
\end{align}
where both logarithms can now be Taylor expanded as $\ln(1+x) = x + O(x^2)$. Keeping the linear terms gives us
\begin{align}
    \Delta \SN &= \sum_{m} -p^m \frac{\delta^m}{p^m} - \delta^m\left(\ln (p^m) + \frac{\delta^m}{p^m}\right) \\
    & =  \sum_{m} -\delta^m  - \delta^m \ln (p^m) - \frac{(\delta^m)^2}{p^m},
\end{align}
The first term above vanishes because the changes cancel by construction. 
Lastly, we carry out the sum and insert the values of $\delta^m$ from the table
\begin{align}
    \Delta \SN  =& -(-2\epsilon)\ln(p^n) - \frac{(-2\epsilon)^2}{p^n}
    -\epsilon\ln(p^{n+1}) - \frac{\epsilon^2}{p^{n+1}} \nonumber \\
    & -\epsilon\ln(p^{n+2}) - \frac{\epsilon^2}{p^{n+2}} \\
    =& - \epsilon \ln\left(\frac{p^{n+1}p^{n+2}}{(p^n)^2}\right)
       - \epsilon^2 \left( \frac{4}{p^n} + \frac{1}{p^{n+1}} + \frac{1}{p^{n+2}} \right) \\
    =& -\epsilon k - O(\epsilon^2),
\end{align}
where the proportionality factor $k$ is positive under the assumption that the number probability is peaked for particle ${n+1}$ at site $L/2$, that is $p^n \approx p^{n+2} \ll p^{n+1}$, which holds in the configuration illustrated in \cref{fig:perturbation}.

\begin{figure}[th]
    \subfloat{%
            \includegraphics[height=39.0mm]{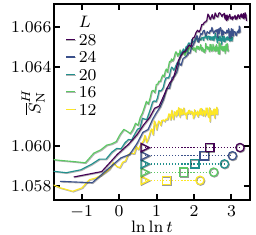}%
            \llap{\parbox[b]{76mm}{\textbf{(a)}\\\rule{0ex}{36mm}}}%
            \label{fig:SH-lnlnt}%
            \begin{picture}(0,0)%
                \put(-74,84.00){\includegraphics[height=10mm]{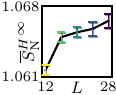}}%
            \end{picture}%
    }%
    \subfloat{%
            \includegraphics[height=39.0mm]{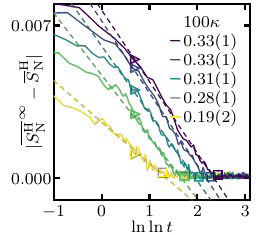}%
            \llap{\parbox[b]{76mm}{\textbf{(b)}\\\rule{0ex}{36mm}}}%
            \label{fig:SH-appr}%
    }%
    \vspace{-3mm}
    \caption{The Hartley number entropy $\SH = S_\mathrm{N}^{(\alpha)}$ disorder averaged for the same dataset as Fig.~2 in the main text, using $\alpha = 10^{-3}$ and probability cutoff $p_\text{c} = 10^{-8}$.
    (a) The growth in double logarithmic time of $\SH$ for various system sizes $L$. The markings denote the time points for growth of and saturation of $\davgSN$ and $\davgSE$. 
    The inset plots the average saturation values as a function of $L$.
    (b) The approach to saturation is fitted to $y = a - \kappa \ln \ln t$ (dashed) for system sizes in (a).
    }
    \label{fig:SH}
\end{figure}

\begin{figure}[th]
    \centering
    \captionsetup[subfigure]{margin=0pt}%
    \setlength{\fboxsep}{0pt}%
    \subfloat{%
            \includegraphics[height=39.0mm]{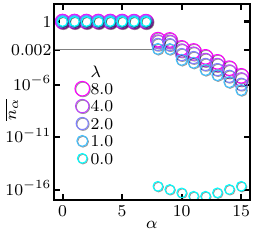}%
            \llap{\parbox[b]{76mm}{\textbf{(a)}\\\rule{0ex}{38mm}}}%
            \label{fig:opdm2}%
    }%
    \subfloat{%
            \includegraphics[height=39.0mm]{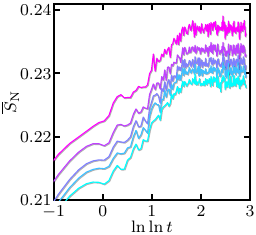}%
            \llap{\parbox[b]{76mm}{\textbf{(b)}\\\rule{0ex}{38mm}}}%
            \label{fig:SN-lambda}%
    }%
    \vspace{-0.35cm}
    \caption{%
    (a) The disorder averaged occupation spectrum $\overline{n_\alpha}$. We calculate eigenvalues $n_\alpha$ of the one-particle density matrix $\rho_{ij} = \bra{\psi'_k} U \sigma^+_i (\prod_{m=i}^{j-1} \sigma_m^z) \sigma_j^- U^\dagger \ket{\psi'_k}$, where $\ket{\psi'_k}$ are product states in the $\ell$-bit basis with $L=16$ sites.
    The $n_\alpha$'s are sorted and averaged over 100 random states $\ket{\psi'_k}$ with $N = L/2$ (zero magnetization).
    Finally, we take the average of 10000 disorder realizations of $U$ to form $\overline{n_\alpha}$.
    (b) We compare the time evolution of disorder averaged number entropy $\davgSN$ varying $\lambda$ for each realization in (a).
    }%
\end{figure}

\begin{figure}[ht]
    \centering
    \captionsetup[subfigure]{margin=0pt}%
    \setlength{\fboxsep}{0pt}%
    \subfloat{%
            \includegraphics[width=0.49\linewidth]{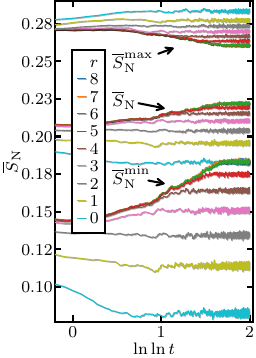}%
            \llap{\parbox[b]{75mm}{\textbf{(a)}\\\rule{0ex}{57mm}}}%
            \label{fig:SN-minmax-lnlnt}%
    }%
    \subfloat{%
            \includegraphics[width=0.49\linewidth]{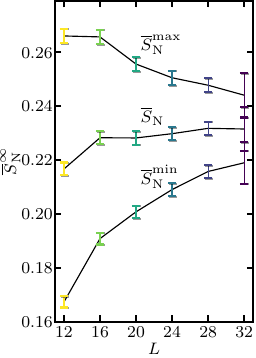}%
            \llap{\parbox[b]{75mm}{\textbf{(b)}\\\rule{0ex}{57mm}}}%
            \label{fig:SN-minmax-L}%
    }
    \vspace{-0.35cm}
    \caption{The average, maxima and minima of $\SN$ following a quench with fixed time step $\delta t = 0.25$.
    We obtain the maxima $\davgSN^{\max}(t)$ (and $\davgSN^{\min}$ similarly) by taking the first peak (valley) found in a sliding time window $\SN(t \leq t' \leq t+10\delta t)$ before disorder averaging. 
    In (a) we set $L=16$ and use 10000 disorder realizations of $H',U$-pairs to time evolve Néel initial states to 6400 time points between $0.25 \leq t \leq 1600$, varying only the maximum range $r = \max(|i-j|)$ in the $\ell$-bit interactions $J_{ij} \sim e^{-|i-j|/\xi_J}$~[see Eq.~(6) in the main text] while $J_{ijk}=0$ .
    In (b) we fix $r=L$ and calculate the saturation values of $\davgSN^{\max}(L)$ and $\davgSN^{\min}(L)$ in the limit $t \rightarrow \infty$. We select a saturated time interval consisting of 40 time points starting at $t = 10^{13}$ and  use the same parameters as in (a) otherwise. 
   }
    \label{fig:SN-minmax}
\end{figure}

\begin{figure*}
    \raggedright
    \setlength{\fboxsep}{0pt}%
    \captionsetup[subfigure]{margin=0pt}%
    \subfloat{%
            \includegraphics[width=59.5mm]{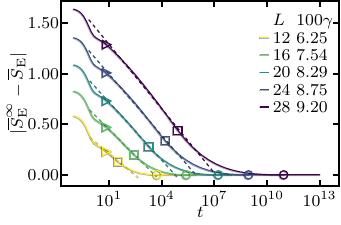}%
            \label{fig:SE-appr}%
    }%
    \llap{\parbox[b]{102mm}{\textbf{(a)}\\\rule{0ex}{38mm}}}%
    \subfloat{%
            \includegraphics[width=59.5mm]{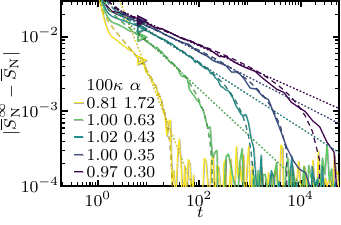}%
            \label{fig:SN-appr-powerfit}%
    }%
    \llap{\parbox[b]{102mm}{\textbf{(b)}\\\rule{0ex}{38mm}}}%
    \subfloat{%
            \includegraphics[width=59.5mm,height=39.5mm]{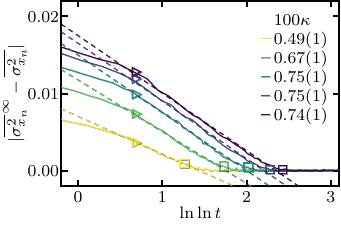}%
            \label{fig:VarX-appr}%
    }%
    \llap{\parbox[b]{102mm}{\textbf{(c)}\\\rule{0ex}{38mm}}}%
    \hfill%

    \vspace{-3mm}%
    \subfloat{%
            \includegraphics[width=59.5mm,height=39.5mm]{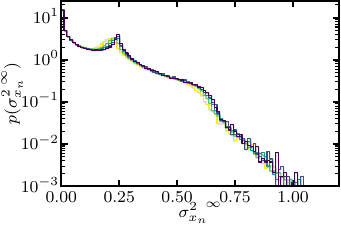}%
            \label{fig:VarX-dist}%
    }%
    \llap{\parbox[b]{102mm}{\textbf{(d)}\\\rule{0ex}{38mm}}}%
    \subfloat{%
            \includegraphics[width=59.5mm,height=39.5mm]{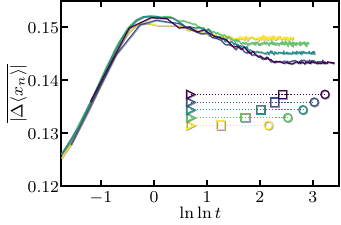}%
            \label{fig:PosX-lnlnt}%
    }%
    \llap{\parbox[b]{102mm}{\textbf{(e)}\\\rule{0ex}{38.3mm}}}%
    \subfloat{%
            \includegraphics[width=59.5mm,height=39.5mm]{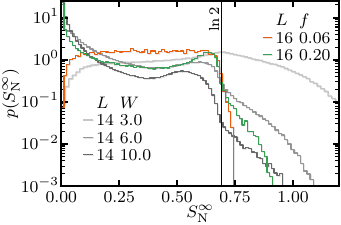}%
            \llap{\parbox[b]{7.8cm}{Ref.~\onlinecite{Luitz2020}:\\\rule{0ex}{23mm}}}%
            \label{fig:SN-dist-weights-off}%
    }%
    \llap{\parbox[b]{102mm}{\textbf{(f)}\\\rule{0ex}{38mm}}}%
    \hfill%
    \vspace{-3mm}
    \caption{All panels are generated from the same dataset as Fig.~(2) in the main text, except for the orange line in panel (f). 
    All fits are taken in the time interval $t_\whtrhd < t < t_\whtsqr$.
    (a) The approach to saturation for $\davgSE$ is fitted to $y = a - \gamma \ln t$ (dashed).
    (b) The approach to saturation for $\davgSN$ is fitted to both $y = a-\kappa \ln \ln t$ (dashed) and $y' = ct ^{-\alpha}$ (dotted).
    (c) The approach to saturation for $\davgVarX$ is fitted to $y = a - \kappa \ln \ln t$ (dashed)
    (d) We use 100 bins in $0 \leq \VarX^\infty \leq 1.2$ for the histogram.
    (e) The disorder-averaged offset of the central quasiparticle position expectation value away from its initial position. 
    (f) The distribution $p(\SN^\infty)$ produced with and without weights $w_i$, at constant $\xi_\tau \approx 0.5$. The case without weights (orange, $w_i=1$) includes $10000$ disorder realizations. Both are compared to results from Ref.~\onlinecite{Luitz2020} (gray) as in the main text.
    }
    \label{fig:additional-data}
\end{figure*}

\section{Hartley number entropy}

The half-chain Hartley number entropy $\SH$ has been proposed as a sensitive probe for large particle number fluctuations that occur with low probability~\cite{Kiefer-Emmanouilidis2021,Kiefer-Emmanouilidis2022}. 
It is defined as the Rényi entropy of the number probability $S^{(\alpha)}_\mathrm{N} = (1-\alpha)^{-1}\ln \sum_n p_{L/2}(n)^\alpha$ in the limit $\alpha\rightarrow 0$, or simply the logarithm of the number of nonzero entries in $p_{L/2}(n)$.
We set $\alpha = 10^{-3}$ and a cutoff $p_\text{c}=10^{-8}$ for the smallest value used in $p_{L/2}(n)$.
The Hartley number entropy is thus presented in \cref{fig:SH}.
Similarly to $\davgSN$, the growth is consistent with $\davgSH \sim \ln \ln t$ during the same time scale, with saturation values that grow for the available system sizes.
The amplitude of the growth depends sensitively on $p_c$, which has a nontrivial interplay with the truncation threshold that we set for the singular value decompositions.

\section{Effect of many-body content in \texorpdfstring{$U$ and $\ell$-bit}{} interaction range}
In microscopic models of many-body localization, slow entropic growth is known to emerge in the presence of local interactions, both for $\SE$~\cite{Bardarson.2012} and $\SN$~\cite{Kiefer-Emmanouilidis2022}. 
When generating the $\ell$-bit basis, these local interactions give rise to long-range higher order terms $\sim \sigma_i^z\sigma_j^z\cdots$ in both $H'$ and $U$.
While in microscopic models $H'$ and $U$ are correlated, in our model they are independent.
Namely, in $H'$ there are random couplings $J_{ij}\tau_i^z\tau_j^z$ and $J_{ijk}\tau_i^z \tau_j^z \tau_k^z$, while in $U$ there are random many-body terms $\frac{\theta_3}{2}\lambda \sigma_{i}^z \sigma_{i+1}^z$ in each gate [see Eqs.~(6-7) and (10) in the main text].
By varying these terms, we can enable the many-body physics selectively and thus attribute the source of entropic growth to either $H'$ or $U$.
The many-body content of $U$ can be characterized by the occupation spectrum of the one-particle density matrix $\rho$~\cite{Bera.2015, Bera.2017}. 
For a given many-body energy eigenstate $\ket{\psi_k}$, one defines $\rho_{ij}=\bra{\psi_k} c_i^\dagger c_j \ket{\psi_k}$, where $c_i^\dagger$ and $c_j$ are the fermionic creation and annihilation operators on sites $i$ and $j$.
The diagonalization $\rho_{ij}\ket{\phi_\alpha} = n_{\alpha} \ket{\phi_\alpha}$ yields a basis of single-particle orbitals $\ket{\phi_\alpha}$ (with $\alpha = 0\dots L-1$), and the eigenvalues $n_\alpha$ (sorted $n_0 \geq n_1 \dots \geq n_{L-1}$)  form the occupation spectrum that adds up to the total number of particles in the system, with $\sum_\alpha n_\alpha = \text{Tr}(\rho) = N$. 
Since in the noninteracting limit the many-body eigenstates are just Slater determinants, the occupation spectrum is a step function $n_0 \dots n_{N-1} = 1$  and $n_{N} \dots n_{L-1} = 0$, such that the gap $\Delta n = n_N - n_{N-1} = 1$.
With interactions, however, the gap reduces to $0 \leq \Delta n < 1$ where $\Delta n \lesssim 1$ in the many-body localized phase, since the eigenstates are dressed orbitals (almost Slater determinants), while $\Delta n \rightarrow 0$ is expected in the ergodic phase in the thermodynamic limit.
In our case, the eigenstates are $\ket{\psi_k} = U^\dagger \ket{\psi'_k}$, where $\ket{\psi'_k}$ are product states (in the $\ell$-bit basis), while the fermionic operators are written in terms of physical spins as $\sigma^+_i (\prod_{m=i}^{j-1} \sigma_m^z ) \sigma_j^-$.
In \cref{fig:opdm2} we present the disorder average $\overline {n_\alpha}$ for different values of $\lambda$, which varies the interaction strength in $U$.
Indeed, when $\lambda = 0$ we recover a step function (up to double precision), and the gap deviates slightly from unity when $\lambda \geq 1$.
Note however that $\davgSN$ in \cref{fig:SN-lambda} (and $\davgSE$, not shown here) is not affected qualitatively by varying $\lambda$: both the initial value and saturation value increases, seemingly linearly with $\lambda$, but the growth remains consistent with $\ln \ln t$ with the same slope and time span.

Next, we reset $\lambda = 1$ and vary instead the upper limit $r$ of the range of pairwise interactions in $H'$, such that $J_{ij} = 0$ whenever $|i-j| > r$ and $J_{ijk}=0$, for system size $L=16$.
The result is presented in \cref{fig:SN-minmax-lnlnt}.
Approaching the limit $r \rightarrow 0$, we see that the growth $\davgSN \sim \ln \ln t$ vanishes, (as does $\davgSE \sim \ln t$, not shown here). 
Meanwhile, the growth appears when $r \geq 3$, and saturates after $r \geq 6$.
Thus, we conclude that the entropic growth is driven by the exponentially decaying couplings in $H'$, and not by the many-body content in $U$.

In \cref{fig:SN-minmax} we also study the extrema $\davgSN^{\min}$ and $\davgSN^{\max}$, which are the disorder averaged dips and peaks in $\SN(t)$, respectively, that may provide an alternative perspective on the origin of $\davgSN$ growth.
The idea is that extrema in $\SN(t)$ correspond to events where the uncertainty in the half-chain particle number $n$ is extremal, such as when quasiparticles randomly oscillate away from, or across, the half-chain boundary.
\cref{fig:SN-minmax-lnlnt} shows that both $\davgSN^{\max}$ and $\davgSN^{\min}$ remain constant in the noninteracting limit $r = 0$, and begin to vary when $r > 0$.
Notably, their behavior is reversed: while $\davgSN^{\min}$ increases (seemingly as $\sim \ln\ln t$), $\davgSN^{\max}$ decreases, albeit to a lesser extent.
In \cref{fig:SN-minmax-L}, we reset $r=L$ and plot the saturation values of $\davgSN^{\max}$ and $\davgSN^{\min}$. 
Again, $\davgSN^{\min}$ increases while $\davgSN^{\max}$ decreases, although now as a function of $L$.
We also see that the amplitude $\davgSN^{\max} - \davgSN^{\min}$ is clearly decreasing, which we take to be consistent with the decreasing standard deviation in $\SN(L)$ seen previously in Ref.~\onlinecite{Luitz2020}.
\section{Additional data}
Aided by \cref{fig:additional-data}, we elucidate some remarks made in the main text. 
In the results section, we have argued that the onset of saturation for $\davgSE$ coincides roughly with the time $t_\whtsqr$ for the full saturation of $\davgSN$. 
Some support for this claim is found in the approach to saturation of $\davgSE$ shown in \cref{fig:SE-appr}, which departs from the linear fit $y\sim\gamma \ln t$ at approximately the same time point as $t_\whtsqr$. 

In \cref{fig:SN-appr-powerfit}, we address the suitability of a power-law fit $\sim t^\alpha$ (dotted) to our data.
By plotting the approach to saturation of $\davgSN$ in log-log axes, it is clear that $\ln \ln t$ (dashed) is a better functional form to describe the transient regime.

The approach to saturation for $\davgVarX$,  shown in \cref{fig:VarX-appr}, demonstrates that the growth $\davgVarX\sim \ln \ln t$ occurs within the same time interval as for $\davgSN \sim \ln \ln t$, namely during $t_\whtrhd < t < t_\whtsqr$, albeit with a smaller slope $\kappa \approx 0.0075$. 

In \cref{fig:VarX-dist}, we show the distribution of saturation values $p(\VarX^\infty)$ for increasing system size $L$. 
Both the peak near $\VarX \approx 0.2$ and the “knee” near $\VarX \approx 0.6$ show a subtle drift with $L$ towards larger values, which results in a slowly increasing disorder average. This resembles the drift seen for $p(\SN^\infty)$ in Fig.~(2d) in the main text, and similarly, it is not clear whether this drift saturates with $L$.

We have argued in the main text that, while the quasiparticle width $\davgVarX$ grows ultraslowly in time, their position expectation values return to the positions given by the initial state.
In \cref{fig:PosX-lnlnt}, we plot the disorder-averaged distance between the central quasiparticle position expectation value and its initial position.
After a quick shift during the early ballistic time scale, the position slowly drifts back toward the initial position in the time scale $t_\whtrhd < t < t_\whtsqr$.
Since the position does not drift away from the initial position over large time scales, it cannot be responsible for the growth of $\davgSN$.

Finally, we provide an example in \cref{fig:SN-dist-weights-off} comparing the distribution of saturation values $p(\SN^\infty)$ with weights on (green, $w_i=e^{-2|h_i-h_{i+1}|}$, see main text Eq.~[9]) and off (orange, $w_i=1$). Note that $f = 0.06$ in the latter, in order to obtain matching localization lengths $\xi_\tau\approx 0.5$ in both cases. At low values $\SN^\infty \approx 0$, the non-weighted case resembles the thermal (low $W$) case from Ref.~\onlinecite{Luitz2020}. In addition, it shows no discernible peak near $\ln 2$, and lacks the tail that follows $\SN > \ln 2$ in the other cases. All these discrepancies motivate the use of weights in the circuit, although we have verified that the growth $\SN\sim\ln\ln t$ is present in both cases.

\bibliography{references.bib}